\documentclass{llncs}
\usepackage[pdftex]{graphicx}
\usepackage[pdftex]{color}
\usepackage{lmodern}
\usepackage{amssymb,bbold}
\usepackage{amssymb} % that is for real numbers etc.
\usepackage{mathtools}
\usepackage{amsfonts}
\usepackage[shortlabels]{enumitem}
\usepackage{natbib}
\usepackage[textfont=it]{caption}
\usepackage[textfont=it]{subcaption}
\newtheorem{Def}{Definition}

\newtheorem{Ex}[Def]{Example}
\newtheorem{Rm}[Def]{Remark}

\newcommand{\bE}{\mathbb{E}}
\newcommand{\bN}{\mathbb{N}}
\newcommand{\bP}{\mathbb{P}}
\newcommand{\bR}{\mathbb{R}}
\newcommand{\bZ}{\mathbb{Z}}
\newcommand{\cF}{\mathcal{F}}
\newcommand{\cM}{\mathcal{M}}
\newcommand{\cS}{\mathcal{S}}
\newcommand{\dist}{\textnormal{dist}}
\DeclareMathOperator*{\argmin}{\textnormal{argmin}}
\begin{document}
% \pagenumbering{gobble} 
\title{Diffusion Means and Heat Kernel on Manifolds} 
\titlerunning{Diffusion Means} 
% \markboth{Dimension Reduction on Polyspheres with Application to Skeletal Representations}{Dimension Reduction on Polyspheres with Application to Skeletal Representations}
\thispagestyle{empty}
% \begin{flushleft}
% \LARGE\bfseries Dimension Reduction on Polyspheres with  Application to Skeletal Representations\\[1cm]
% \end{flushleft}
\author{Pernille Hansen\inst{1}, Benjamin Eltzner\inst{2} \and  Stefan Sommer\inst{1}}
\authorrunning{Hansen, Eltzner, Sommer}
\institute{$^1$University of Copenhagen, Denmark, Department of Computer Science \\
  $^2$Georg-August-Universit\"at at G\"ottingen, Germany, Felix-Bernstein-Institute for Mathematical Statistics in the Biosciences}

\maketitle
\thispagestyle{plain}

\begin{abstract}
  We introduce diffusion means as location statistics on manifold data spaces. A diffusion mean is defined as the starting point of an isotropic diffusion with a given diffusivity. They can therefore be defined on all spaces on which a Brownian motion can be defined and numerical calculation of sample diffusion means is possible on a variety of spaces using the heat kernel expansion. We present several classes of spaces, for which the heat kernel is known and sample diffusion means can therefore be calculated. As an example, we investigate a classic data set from directional statistics, for which the sample Fr\'echet mean exhibits finite sample smeariness.
\end{abstract}

\section{Introduction}

In order to analyze data which are represented not on a vector space but a more general space $\cM$, where we focus on manifolds here, it is necessary to generalize concepts from Euclidean space to more general spaces. Important examples of data on non-Euclidean spaces include directional data, cf. \cite{MJ00}, and landmark shape spaces, see \cite{small_statistical_1996}.

In the field of general relativity where similar generalizations are required, one usually relies on the \emph{correspondence principle} as a minimal requirement for quantities on curved spaces. It states that an observable in curved space should reduce to the corresponding observable in the flat case. In terms of statistics this means that any generalization $S_\cM$ of a statistic $S$ on Euclidean space to a manifold $\cM$ should reduce to $S$ on Euclidean space.

The mean is the most widely used location statistic and was generalized to metric spaces by \cite{frechet_les_1948}, who defined it as the minimizer of expected squared distance. This definition of the Fr\'echet mean satisfies the correspondence principle. However, as is often the case in physics, it is by far not the only parameter that has this property. In order to judge which potential other generalizations of the mean are meaningful, we recall that one of the reasons the mean is so widely used and useful is that it is an estimator of (one of) the model parameter(s) in many parametric families of probability distributions including Normal, Poisson, Exponential, Bernoulli and Binomial distribution. It is therefore useful to link potential generalizations of the mean to parametric families on general data spaces.

Few parametric families of probability distributions have been widely generalized to non-Euclidean data spaces or even to $\bR^m$ with $m > 1$. One class of distributions which can be widely generalized are isotropic normal distributions, as these can be defined as the distribution of a Brownian motion after unit time. Since Brownian motion can be defined on any manifold, this definition is very general. The thus defined probability distribution have one location parameter $\mu^t \in \cM$ and a spread parameter $t \in \bR^+$. Its clear interpretation in terms of the generalized normal distribution family makes the \emph{diffusion mean} $\mu^t$ an interesting contender for a generalization of the Euclidean mean.

The location parameter of Brownian motion can be defined using the heat kernels $p(x,y,t)$, which are the transition densities of Brownian motions. For a fixed $t>0$, we define the \emph{diffusion $t$-mean set} as the minima 
\begin{equation} \label{eq:diff-mean}
  E^t(X) = \argmin_{\mu^t \in P}\limits \bE[-\ln p(X,y,t)].
\end{equation}
The logarithmic heat kernel is naturally connected to geodesic distance due to the limit $\lim_{t \to 0} \limits -2t\ln p(x,y,t) = \dist(x,y)^2$, cf. \cite{hsu_stochastic_2002}, which means that the Fr\'echet mean can be interpreted as the $\mu^0$ diffusion mean, i.e. the limit for $t \to 0$. In Euclidean space, Equation \eqref{eq:diff-mean} reduces to the MLE of the normal distribution, which means that $\mu^t$ does not depend on $t$. On other data spaces, the two parameters do not decouple in general and $\mu^t$ can be different depending on $t$. Since all diffusion means satisfy the correspondence principle for the Euclidean mean, it is not immediately clear why the Fr\'echet mean should be preferred over diffusion means for finite $t > 0$.

Explicit expressions for the heat kernel are known for several classes of manifolds and for a more general class a recursively defined asymptotic series expansion exists. This means that diffusion means can be numerically determined on many data spaces and their dependence on $t$ can be studied. In this article, we focus on \emph{smeariness of means} as described by \cite{hotz_intrinsic_2015,eltzner_smeary_2018,Eltzner2020}, more precisely finite sample smeariness as described by \cite{HEH19}. A smeary mean satisfies a modified central limit theorem with a slower asymptotic rate than $n^{-1/2}$. This affects samples drawn from such a population, whose sample means can exhibit \emph{finite sample smeariness}.

After a brief overview of the relevant concepts, we will give a number of examples of data spaces in which diffusion means can be readily computed. Lastly, we investigate the diffusion means for a directional data set on $S^1$ whose Fr\'echet mean exhibits finite sample smeariness. We find that the diffusion means exhibit less finite sample smeariness with increasing $t$.

\section{Basic Concepts and Definitions}

%\subsection{Differential Geometry}
A Riemannian manifold $(\cM,g)$ is a smooth manifold equipped with inner products $\langle .,.\rangle_x$ on the tangent spaces $T_x\cM$ for each $x\in\cM$ such that $x\mapsto \langle v_x,u_x\rangle_x$ is smooth for all vector fields $u,v\in T\cM$. Curves which are locally length minimizing are called \emph{geodesics}. The \emph{Riemannian distance} between two points is defined by the infimum over the lengths of geodesics connecting the points. If there exist more than one length minimizing geodesic joining $x$ and $y$, we say that $y$ is in the cut of $x$ and define the \emph{cut locus} $C(x)$ as the collection of all such points. 

For every starting point $x \in \cM$ and velocity vector $v \in T_x\cM$ there is a unique geodesic $\gamma_{x,v}$ and the exponential map $\exp_x:T_x\cM\to \cM$ maps $v$ to the point reached in unit time $\gamma_{x,v}(1)$. The exponential map is a diffeomorphism on a subset $D(x)\subset T_x\cM$ such that its image coincides with $\cM\backslash C(x)$, and the logarithm $\log_x:\cM\backslash C(x) \to T_x\cM$ is defined as the inverse map. 

The heat kernel is the fundamental solution to the heat equation
\begin{align*}
  \frac{d}{dt} p(x,y,t) = \frac{1}{2} \Delta_x p(x,y,t)
\end{align*}
where $\Delta$ is the Laplace Beltrami operator on $\cM$ and it is also the transition density of Brownian motions on $\cM$. A Riemannian manifold is stochastically complete if there exists a minimal solution $p$ satisfying $\int_\cM p(x,y,t)dy = 1$ for all $x\in \cM$ and $t>0$. The minimal solution is strictly positive, smooth and symmetric in its first two arguments.

Let $(\Omega, \mathcal F, \mathbb P)$ be the underlying probability space of the random variable $X$ on the stochastically complete manifold $\cM$ with minimal heat kernel $p$. We define the log-likelihood function $L^t:\cM\to \mathbb R$ to be
\begin{equation}
	L^t(y) = \bE[-\ln p(X,y,t)]
\end{equation}
for $t>0$.  This gives rise to the diffusion means as the global minima of the log-likelihood function, i.e. the points maximizing the log-likelihood of a Brownian motion.
\begin{Def}
	With the underlying probability space $(\Omega, \cF,\bP)$, let $X$ be a random variable on $\cM$ and fix $t>0$. The \emph{diffusion $t$-mean set} $E^t(X)$ of $X$ is the set of global minima of the log-likelihood function $L^t$, i.e. 
	\begin{equation*}
	E^t(X) = \argmin_{y\in \cM} \limits \bE[-\ln(p(X,y,t))].
	\end{equation*}
	If $E^t(X)$ contains a single point $\mu^t$, we say that $\mu^t$ is the \emph{diffusion $t$-mean} of $X$. 
\end{Def}
We consider the asymptotic behavior and smeariness of the following estimator. 
\begin{Def} For samples $X_1,..., X_n\overset{\text{i.i.d.}}{\sim} X$ on $\cM$ we define the \emph{sample log-likelihood function} $L^t_n:\cM \to \bR$, 
	\begin{equation*}
	L^t_n(y) = - \frac{1}{n} \ln\Big(\prod_{i=1}^{n}p(X_i,y,t) \Big) = -\frac{1}{n} \sum_{i=1}^n \ln p(X_i,y,t)
	\end{equation*}
	for every $n\in\bN$ and the \emph{sample diffusion $t$-mean sets} 
	$E_{t,n} = \argmin_{y\in \cM} \limits L^t_n(y).$
\end{Def}

\begin{Def}[Smeariness of Diffusion Means] \label{def:smeary}
  Consider a random variable $X$ on $\mathcal{M}$, an assume that there is $\zeta > 0$ such that for every $x \in B_\zeta (0) \setminus \{0\}$ one has $L^t(\exp_\mu(x)) > L^t(\mu^t)$. Suppose that for fixed constants $C_X > 0$ and $2 < \kappa \in \mathbb{R}$ we have for every sufficiently small $\delta > 0$
  \begin{align*}
    \sup_{x \in T_{\mu^t} \cM, \, \|x\| < \delta} \limits \left| L^t(\exp_\mu(x)) - L^t(\mu^t) \right| &\ge C_X \delta^\kappa \, .
  \end{align*}
  Then we say that the diffusion mean $\mu^t$ of $X$ is \emph{smeary}.
\end{Def}

\begin{Def}[Finite Sample Smeary Mean]
  For the population mean $\mu^t \in \cM$ and its corresponding sample estimator $\widehat{\mu}^t_n$ let
  \begin{align*}
    \mathfrak{m}^t_n := \frac{n \bE[d^2(\widehat{\mu}^t_n, \mu^t)]}{\bE[d^2(X, \mu^t)]}
  \end{align*}
  be the \emph{variance ratio of $\mu^t$}. Then $\mu^t$ is called \emph{finite sample smeary}, if
  \begin{align*}
    S^t_\textnormal{FSS} := \sup_{n \in \bN} \limits \mathfrak{m}^t_n > 1 \qquad \textnormal{and} \qquad S^t_\textnormal{FSS} < \infty \, .
  \end{align*}
  The latter requirement distinguishes finite sample smeariness from smeariness, where $\lim_{n\to\infty} \limits \mathfrak{m}^t_n = \infty$.
\end{Def}

A consistent estimator for $\mathfrak{m}^t_n$ can be given using the n-out-of-n bootstrap
\begin{align*}
  \widehat{\mathfrak{m}}^t_n := \frac{n \frac{1}{B}\sum_{b=1}^B d^2(\mu_{n,n}^{t,*b},\widehat{\mu}^t_n)}{\frac{1}{n}\sum_{j=1}^n d^2(\widehat{\mu}^t_n,X_j)}\, .
\end{align*}
This estimator is used in the application below.

\section{Examples of known Heat Kernels}

\begin{Ex}
  The heat kernel $p$ on the Euclidean space $\bR^m$ is given by the function
  \begin{equation*}
  p(x,y,t) = \left(\frac{1}{(4\pi t)^{m/2}}\right) e^{\frac{-|x-y|^2}{4t}}. 
  \end{equation*}
  for $x,y\in \bR^m$ and $t>0$. The diffusion $t$-means of a random variable $X$ does not depend on $t$ and coincide with the expected value $\bE[X]$ since  
  \begin{align*}
  \argmin_{y\in \bR^m} L^t(y)%&=\argmin_{y\in \bR^m} \int_{\bR^m} \frac{2}{m}\ln(4\pi t) - \left(\frac{|x-y|^2}{4t}\right)d\bP_X(x) \\
  %&= \argmin_{y\in \bR^m}  \int_{\bR^m} |x-y|^2d\bP_X(x) \\
  &=  \argmin_{y\in \bR^m} \bE[(X-y)^2]
  \end{align*}
  Thus, $\mu^t = \bE[X]$ for all $t>0$. 
\end{Ex}
\begin{Ex}
  The heat kernel on the circle $\cS^1$ is given by the wrapped Gaussian
  \begin{equation*}
  p(x,y,t) = \frac{1}{\sqrt{4\pi t}} \Big( \sum_{k\in \bZ} \exp\Big(\frac{-(x-y + 2\pi k)^2}{4t}  \Big) \Big)
  \end{equation*}
  for $x,y\in \bR / \bZ\cong \cS^1$ and $t>0$, and the log-likelihood function for the random variable $X:\Omega \to \cS^1$ becomes
  \begin{align*}
  L^t(y) &=  - \ln \Big(\sqrt{4\pi t} \Big) + \int_{\cS^1} \ln \Big( \sum_{k\in \bZ}\exp\Big(\frac{-(x-y + 2\pi k)^2}{4t}  \Big) \Big) d\bP_X(x)% \\
  %&=  \int_{\cS^1}-\ln(\sqrt{4\pi t})+ \frac{(x-y)^2}{4t} - \ln \Big(\sum_{k\in \bZ}\exp\Big(\frac{-(2\pi k)^2}{4t}\Big)\exp\Big(\frac{-4\pi k (x-y)}{4t}  \Big) \Big) d\bP_X(x).
  \end{align*}
  Notably, even on this simple space, the $t$-dependence in the exponentials is not a simple prefactor and $\mu^t$ is therefore explicitly dependent on $t$.
\end{Ex}
\begin{Ex}
  The heat kernel on the spheres $\cS^{m}$ for $m\geq 2$ can be expressed as the uniformly and absolutely convergent series, see \cite[Theorem 1]{zhao_exact_2018},
  \begin{equation*}%\label{sphereheat}
  p(x,y,t) = \sum_{l=0}^\infty e^{-l(l+m-1)t}\frac{2l+m-1}{m-1} \frac{1}{A_{\cS}^{m}} C_l^{(m-1)/2}(\langle x,y\rangle_ {\bR^{m+1}} )
  \end{equation*}
  for $x,y\in \cS^{m}$ and $t>0$, where $C_l^\alpha$ are the Gegenbauer polynomials and $A_{\cS}^{m} = \frac{2\pi^{(m+1)/2}}{\Gamma((m+1)/2)}$ the surface area of $\cS^{m}$. For $m=2$, the Gegenbauer polynomials $C^{1/2}_l$ coincide with the Legendre polynomials $P^0_l$ and the heat kernel on $\cS^2$ is 
  \begin{equation*}
  p(x,y,t) = \sum_{l=0}^\infty e^{-l(l+1)t}\frac{2l+1}{4\pi} P^0_l(\langle x,y\rangle_ {\bR^3} ).
  \end{equation*}
  Again, $\mu^t$ is explicitly dependent on $t$ on these spaces.
\end{Ex}
\begin{Ex}
  The heat kernel on the hyperbolic space $\mathbb{H}^m$ can be expressed by the following formulas, see \cite{grigoryan_heat_1998}, for $n\geq 1$. For odd $m =2k+1$, the heat kernel is given by
  \begin{equation*}
  p(x,y,t) = \frac{(-1)^k}{2^k\pi^k}\frac{1}{\sqrt{4\pi t}}\frac{\rho}{\sinh\rho}\left(\frac{1}{\sinh\rho} \frac{\partial}{\partial\rho} \right)^k e^{-k^2t-\frac{\rho^2}{4t}}
  \end{equation*}
  where $\rho = \dist_{\mathbb{H}^m}(x,y)$ and for even $m=2k+2$, it is given by 
  \begin{equation*}
    p(x,y,t) = \frac{(-1)^k}{2^{k+\frac52}\pi^{k+\frac32}}t^{-\frac32}\frac{\rho}{\sinh\rho}\left(\frac{1}{\sinh\rho} \frac{\partial}{\partial\rho} \right)^k  \int_\rho^\infty \frac{s \exp\left(-\frac{s^2}{4t}\right)}{(\cosh s-\cosh \rho)^\frac12}ds.
  \end{equation*}
  Again, $\mu^t$ is explicitly dependent on $t$ on these spaces.
\end{Ex}
\begin{Ex} The fundamental solution to the heat equation on a Lie group $G$ of dimension $m$ is 
  \begin{equation*}
  p(x,e,t) = (2\pi t)^{-m/2} \prod_{\alpha\in\Sigma^+} \frac{i\alpha(H)}{2\sin(i\alpha(H)/2)}\exp\left(\frac{\|H\|^2}{2t}+\frac{\|\rho\|^2 t}{2}\right) \cdot E_x(X_\tau >t)
  \end{equation*}
  where $e$ is the neutral element, $x = \exp(Ad(g)H)\in G\setminus\mathcal{C}(e)$ for some $g\in G$, $\Sigma^+$ is the set of positive roots, $\rho = \sum_{\alpha\in\Sigma^+} \alpha$, and $\tau$ is the hitting time of $\mathcal{C}(e)$ by $(x_s)_{0\leq s\leq t}$. Lie groups are relevant data spaces for many application, e.g. in the modeling of joint movement for robotics, prosthetic development and medicine.
\end{Ex}
The symmetry in the example above is quite noticeable and in fact when $t>0$ and $x\in \cM$ where $\cM = \bR^m, \cS^m$ or $\mathbb{H}^m$, the heat kernel only depends on the geodesic distance, see \cite{alonso-oran_pointwise_2019}. 

\begin{Rm}
  For spaces where a closed form has not been obtained, we can turn to various estimates. We include some of the most well known below. 
  \begin{enumerate}
    \item For complete Riemannian manifolds of dimension $m$, we have the asymptotic expansion, see \cite{hsu_stochastic_2002},
    \begin{equation*}
    p(x,y,t)\sim \left(\frac{1}{2\pi t} \right)^{\frac{m}{2}}e^{\frac{-d(x,y)^2}{2t}}\sum_{n=0}^{\infty} H_n(x,y)t^n
    \end{equation*}
    on compact subset with $x\notin \mathcal{C}(y)$. Here $H_n$ are smooth functions satisfying a recursion formula, see \cite{chavel_eigenvalues_1984}, with $H_0(x,y) = \sqrt{J(\exp_x)(\exp^{-1}_x(y))}$ and $J$ denoting the Jacobian.
    \item Assuming also non-negative Ricci curvature, the heat kernel is bounded from both sides, see \cite{grigoryan_heat_1994,saloff-coste_note_1992},
    \begin{equation*}
    \frac{c_1}{\text{vol}(x,\sqrt{t})}\exp \left(\frac{\dist(x,y)^2}{c_2t}\right) \leq p(x,y,t) \leq \frac{c_3}{\text{vol}(x,\sqrt{t})}\exp \left(\frac{\dist(x,y)^2}{c_4t}\right)
    \end{equation*}
    where $\text{vol}(x,\sqrt{t})$ denotes the volume of the ball around $x$ of radius $\sqrt{t}$ and positive constants $c_i$ for $i = 1,...,4$.
    \item Using bridge sampling, the heat kernel can be estimated by the expectation over guided processes, see \cite{delyon_simulation_2006,JS2021}. %,sorensen_importance_2012
    An example of this is the estimated heat kernel on landmark manifolds by \cite{sommer_bridge_2017}. 
  \end{enumerate}
\end{Rm}

\section{Application to Smeariness in Directional Data}

In this Section we apply diffusion means to the classic data set denoting the compass directions of sea turtles leaving their nest after egg laying, cf. \cite{Stephens1969} and \citet[p. 9]{MJ00}. The data set is clearly bimodal with two antipodal modes. The main mode is in north-northwestern direction, while the smaller mode is exactly in the opposite direction. It was shown in \cite{eltzner_smeary_2018} that the Fr\'echet mean for this data set exhibits pronounced finite sample smeariness.

\begin{figure}[ht!]
  \centering
  \subcaptionbox{Data histogram and Fr\'echet functions}[0.75\textwidth]{\includegraphics[width=0.75\textwidth]{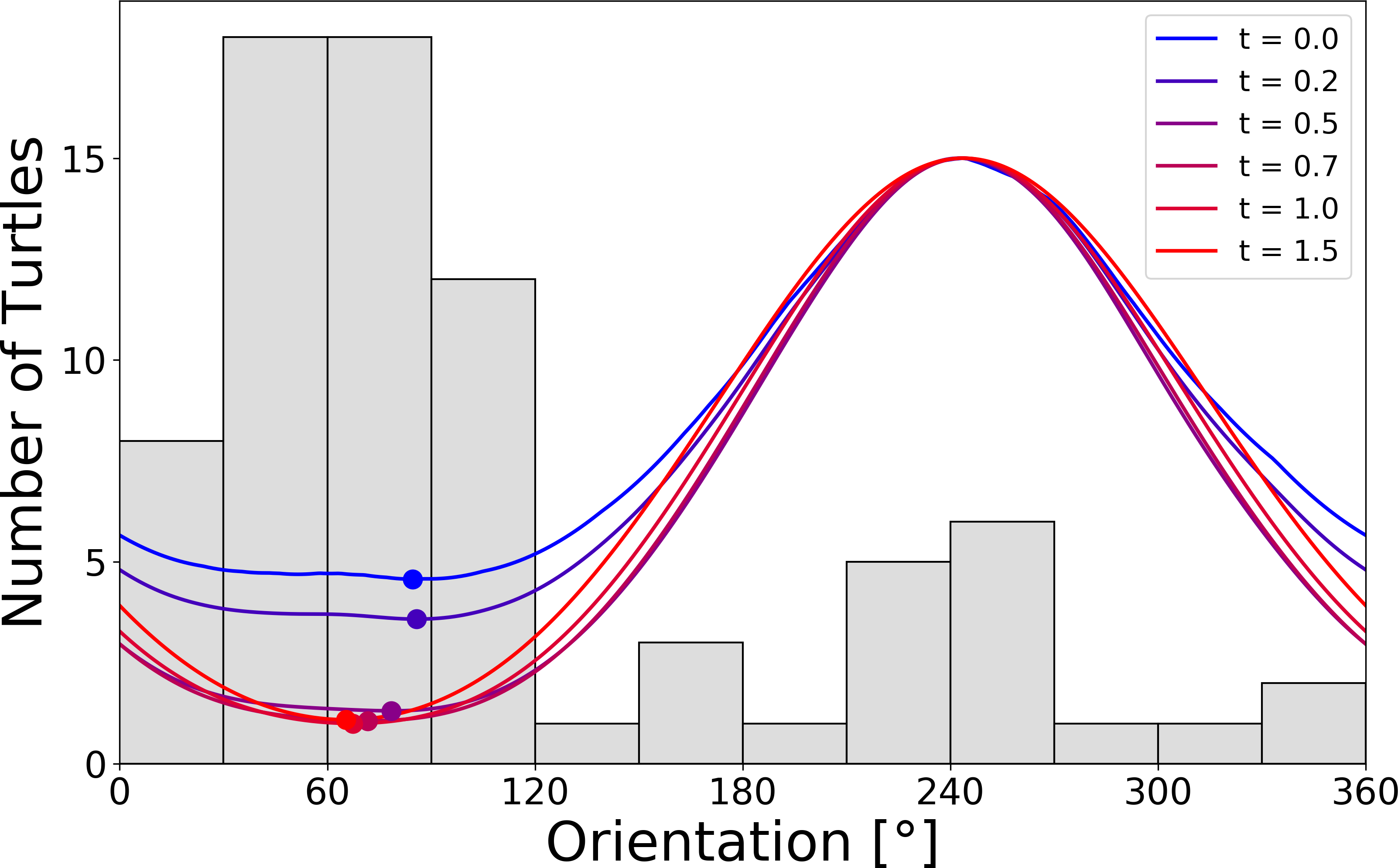}}\\
  \vspace*{\baselineskip}
  \subcaptionbox{Variances of sample means}[0.75\textwidth]{\includegraphics[width=0.75\textwidth]{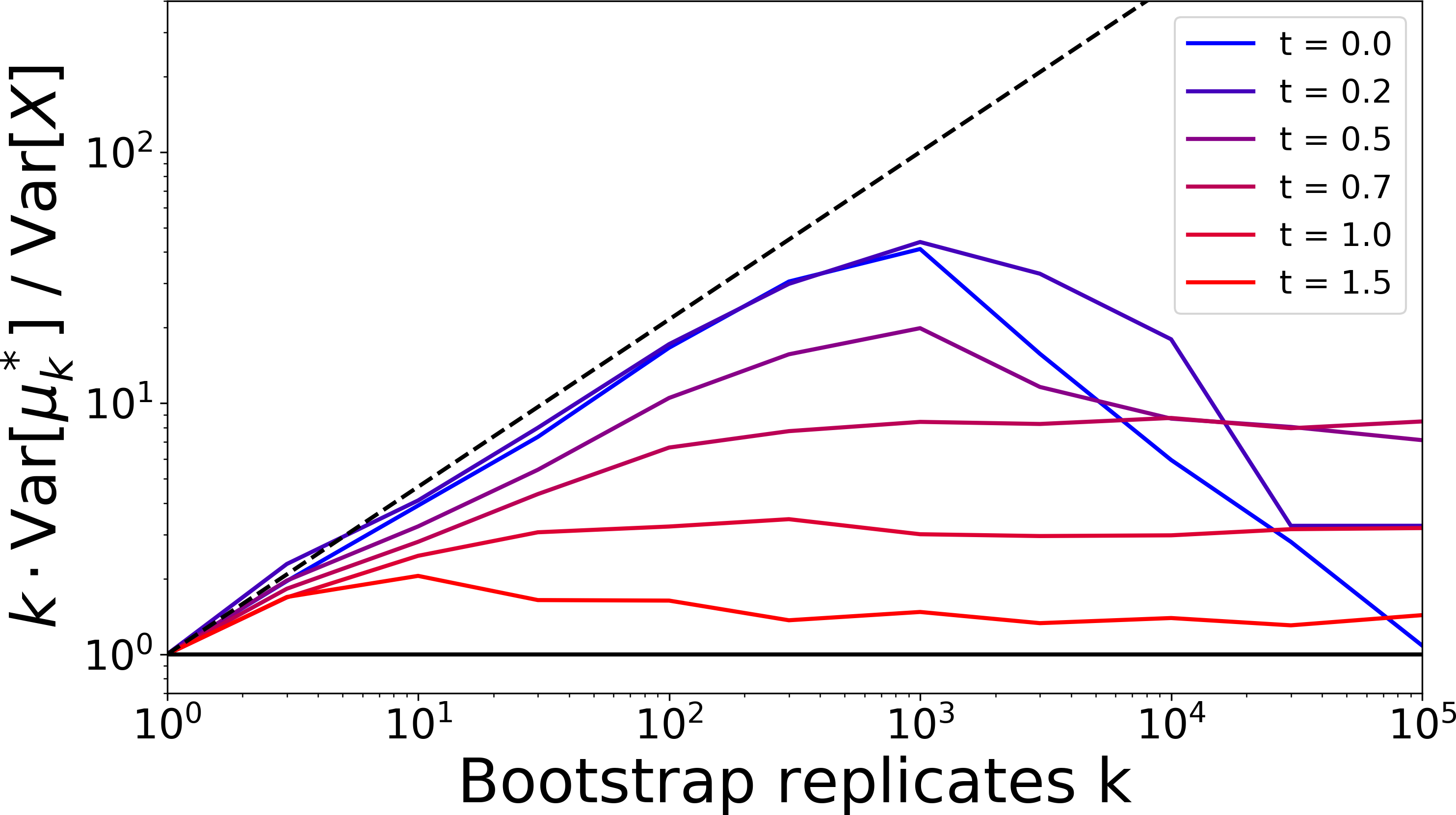}}
  \caption{Diffusion means for nesting sea turtles. Panel~(a) shows the main mode of turtle directions in the east-northeastern direction of the sea shore and a second mode in exactly opposite direction. The sample likelihood functions $L^t_n$ are rescaled to a common scale but their relative minimum values were preserved. With increasing $t$ the minimum of $L^t_n$ approaches the center of the main mode and $L^t_n$ approaches an $x^2$ behavior at the minimum. Panel~(b) shows that, correspondingly, finite sample smeariness decreases in magnitude with increasing $t$. \label{fig:turtles}}
\end{figure}

In Figure \ref{fig:turtles}a, we show that increasing the diffusivity $t$, the likelihood function at the minimum, starting out unusually flat, approaches a parabolic shape. In Figure \ref{fig:turtles}b, we show the corresponding curves of estimated variance ratio $\widehat{\mathfrak{m}}^t_n$, whose maxima can be used as estimators for the magnitude $S^t_\textnormal{FSS}$ of finite sample smeariness. As expected from the visual inspection of the likelihoods, the magnitude of finite sample smeariness decreases with increasing $t$.

This behavior of reducing the effects of smeariness has been found analogously in other applications and simulations. These results suggest that diffusion means can provide a more robust location statistic than the Fr\'echet mean for spread out data on positively curved spaces. This point is reinforced by estimating $t$ and $\mu^t$ jointly, which yields $\widehat{t} = 0.963$. As one can see from the case $t=1$ in Figure \ref{fig:turtles}, this leads to a very low magnitude of finite sample smeariness compared to the Fr\'echet mean.

%\bibliographystyle{Chicago}
%\bibliography{BibPaper}

\end{document}